\documentclass[useAMS,usenatbib]{mn2e}

 \usepackage{times}
 \usepackage{graphicx}

\newbox\grsign \setbox\grsign=\hbox{$>$} \newdimen\grdimen
\grdimen=\ht\grsign
\newbox\simlessbox \newbox\simgreatbox \newbox\simpropbox
\setbox\simgreatbox=\hbox{\raise.5ex\hbox{$>$}\llap
     {\lower.5ex\hbox{$\sim$}}}\ht1=\grdimen\dp1=0pt
\setbox\simlessbox=\hbox{\raise.5ex\hbox{$<$}\llap
     {\lower.5ex\hbox{$\sim$}}}\ht2=\grdimen\dp2=0pt
\setbox\simpropbox=\hbox{\raise.5ex\hbox{$\propto$}\llap
     {\lower.5ex\hbox{$\sim$}}}\ht2=\grdimen\dp2=0pt
\def\simgreat{\mathrel{\copy\simgreatbox}}
\def\simless{\mathrel{\copy\simlessbox}}

\title[Candidate $\alpha$ Per white dwarfs]{Optical spectroscopy of  candidate Alpha Persei white dwarfs}
\author[S.L.Casewell et al.]{S.L.Casewell$^{1}$\thanks{E-mail:
slc25@le.ac.uk},  P. D. Dobbie$^{2}$,  S. Geier$^{3,4}$, N. Lodieu$^{5,6}$ \& N.C. Hambly$^{7}$\\
$^{1}$Department of Physics and Astronomy, University of Leicester, University Road, Leicester LE1 7RH, UK \\
$^{2}$School of Physical Sciences, University of Tasmania, Hobart, TAS 7001, Australia \\
$^{3}$European Southern Observatory, Karl-Schwarzschild-Str. 2, 85748, Garching, Germany\\
$^{4}$Dr.~Karl~Remeis-Observatory \& ECAPWD, Astronomical Institute, Friedrich-Alexander University Erlangen-Nuremberg, Sternwartstr.~7, D 96049 Bamberg, Germany\\
$^{5}$Instituto de Astrof\'isica de Canarias (IAC), C/ V'a L\'actea s/n, E-38200 La Laguna, Tenerife, Spain\\
$^{6}$Departamento de Astrof\'isica, Universidad de La Laguna (ULL), E-38206 La Laguna, Tenerife, Spain\\
$^{7}$Scottish Universities Physics Alliance (SUPA), Institute for Astronomy, School of Physics, University of Edinburgh, Royal Observatory, Blackford Hill, Edinburgh EH9 3HJ}

\begin{document}

\date{Accepted \today{}. Received \today{}; in original form \today{}}

\pagerange{\pageref{firstpage}--\pageref{lastpage}} \pubyear{2014}

\maketitle

\label{firstpage}

\begin{abstract}
As part of an investigation into the high mass end of the initial mass-final mass relation we performed a search for new white dwarf members of the nearby (172.4 pc), young (80-90  Myr) $\alpha$ Persei open star cluster. The photometric and astrometric search using the UKIRT Infrared Deep Sky Survey  and SuperCOSMOS sky surveys discovered 14 new white dwarf candidates. We have obtained medium resolution optical spectra of  the brightest 11 candidates  using the William Herschel Telescope and confirmed that while 7 are DA white dwarfs, 3 are DB white dwarfs and one is an sdOB star, only three have cooling ages within the cluster age, and from their position on the initial mass-final mass relation, it is likely none are cluster members. This result is disappointing, as recent work on the cluster mass function suggests that there should be at least one white dwarf member, even at this young age. It may be that any white dwarf members of $\alpha$ Per are hidden within binary systems, as is the case in the Hyades cluster, however the lack of high mass stars within the cluster also makes this seem unlikely. One alternative is that a significant level of detection incompleteness in the legacy optical image survey
data at this Galactic latitude has caused some white dwarf members to be overlooked. If this is the case, Gaia will find them.

\end{abstract}


\section{Introduction}

The initial mass-final mass relation (IFMR) describes the relationship between the main sequence mass of a star
with M$\simless$10 M$_{\odot}$ and the mass of the white dwarf created after it dies (e.g. \citealt*{iben83}).  
Understanding the form of this relation is important since it provides information on the amount of gas enriched with He, N and other
metals that are returned  to the interstellar medium by the death of low- and intermediate-mass stars.
Additionally, the form of the upper end of the IFMR is relevant to studies of Type II supernovae as it can provide a constraint on the 
minimum mass of star that will experience this fate (e.g. \citealt{siess06}).

The form of a theoretical  IFMR is extremely difficult to predict  due to the many complex processes occurring during the final phases
of stellar evolution (e.g. third dredge-up, thermal pulses, mass loss). This means that robust empirical data are essential for constraining 
its form. However, these data can be challenging to obtain, as it requires determining the main sequence mass of a
star that has long since died.  This difficulty can be alleviated by investigating white dwarf members of open star clusters 
\citep{weidemann77,weidemann00,dobbie06, casewell09, dobbie12}. Here, since the age of the population can be determined from the location of the main sequence
turn-off \citep{king05} or a lithium age if the cluster is sufficiently young, the lifetime and mass of the progenitor star of any white dwarf member can be estimated by calculating 
the difference between the cooling time of the white dwarf and the cluster age.

During the last decade we have seen substantial progress in mapping the IFMR, with several groups exploiting mosaic imagers and telescopes with blue 
sensitive spectrographs to perform detailed studies of cluster white dwarfs (e.g. \citealt{kalirai07,williams09,casewell09,dobbie09}). However, despite clear headway, the IFMR remains very poorly sampled by observations for M$_{\rm init}$$\simgreat $5.5-6M$
_{\odot}$. Indeed, there are only a handful of  white dwarfs here \citep{williams09, dobbie12}. So, while it is extremely important to have a good understanding 
of the top end of the IFMR, its form here remains greatly uncertain.

In a bid to obtain crucial new data in this initial mass regime,
we have used the extensive imaging obtained as part of the UKIRT Infrared Sky Survey Galactic Clusters Survey (UKIDSS GCS: \citealt{lawrence07}) to search the open cluster $\alpha$ Per for candidate white dwarf members. This population 
has several characteristics which suggest it is particularly well suited to this type of investigation yet until now it 
has not been exploited. It is nearby, 172.4 pc \citep{vanleuwen09} and despite residing at low Galactic latitude (b=-6.053, \citealt{kharchenko13}),
 foreground extinction is low, E$_{B-V}$$<$0.1 \citep{prosser92}. Thus
intrinsically faint members will appear comparatively bright and can be studied in detail with spectrographs on modern telescopes in modest
integration times. $\alpha$ Per also has a distinct proper motion ($\mu$$_{\alpha}$ cos $\delta$, $\mu$$_{\delta}$ $\sim$+23, -27 mas\,yr$^{-1}$; \citealt{vanleuwen09})  which helps to distinguish members for the general field population. The cluster age is especially well constrained, $\tau$=90$\pm$10 Myr (\citealt{stauffer99}
: corresponding to M$_{\rm initial}$$\sim$5.5 M$_{\odot}$), via the lithium depletion boundary technique \citep{stauffer99}, helping to minimise uncertainty in the progenitor mass determinations. The main sequence turn off age for this cluster is 50 Myr \citep{mermilliod81}, but for young clusters the lithium age is more reliable, and so in this work we use 90$\pm$10 Myr as the cluster age.

Moreover, $\alpha$ Per is sufficiently old for white dwarfs to have formed, but young 
enough that the oldest of these remain at T$_{\rm eff}$$\simgreat$12500 K. Hydrogen rich atmospheres of white dwarfs are also dominated by radiative 
energy transport in this region, so models are less complicated, and more reliable in this regime \citep{bergeron95}. \newline
\indent Studies of the $\alpha$ Per cluster have located some higher mass members (\citealt{prosser92, prosser98, deacon04}) but because of its youth, most studies have mainly concentrated on discovering new brown dwarf members (e.g. \citealt{barrado02, deacon04, lodieu05, lodieu12}). A detailed mass function of the cluster was calculated in \citet{lodieu12} who studied $\sim$56 square degrees of the cluster and determined that the mass function is similar in shape to that of the Pleiades and can be represented by a log normal with a characteristic mass of 0.34 M$_{\odot}$ and a dispersion of 0.46. This similarity to the Pleiades indicates that $\alpha$ Per may indeed harbour white dwarf members and is constant with 
results from older studies  (e.g. \citealt{sanner01}) which suggest that $\alpha$ Per may be richer than the Pleiades.


%
\section{Sample selection}
\label{data}
\begin{figure}
\begin{center}
\scalebox{0.3}{\includegraphics[bb = 7 7 661 559]{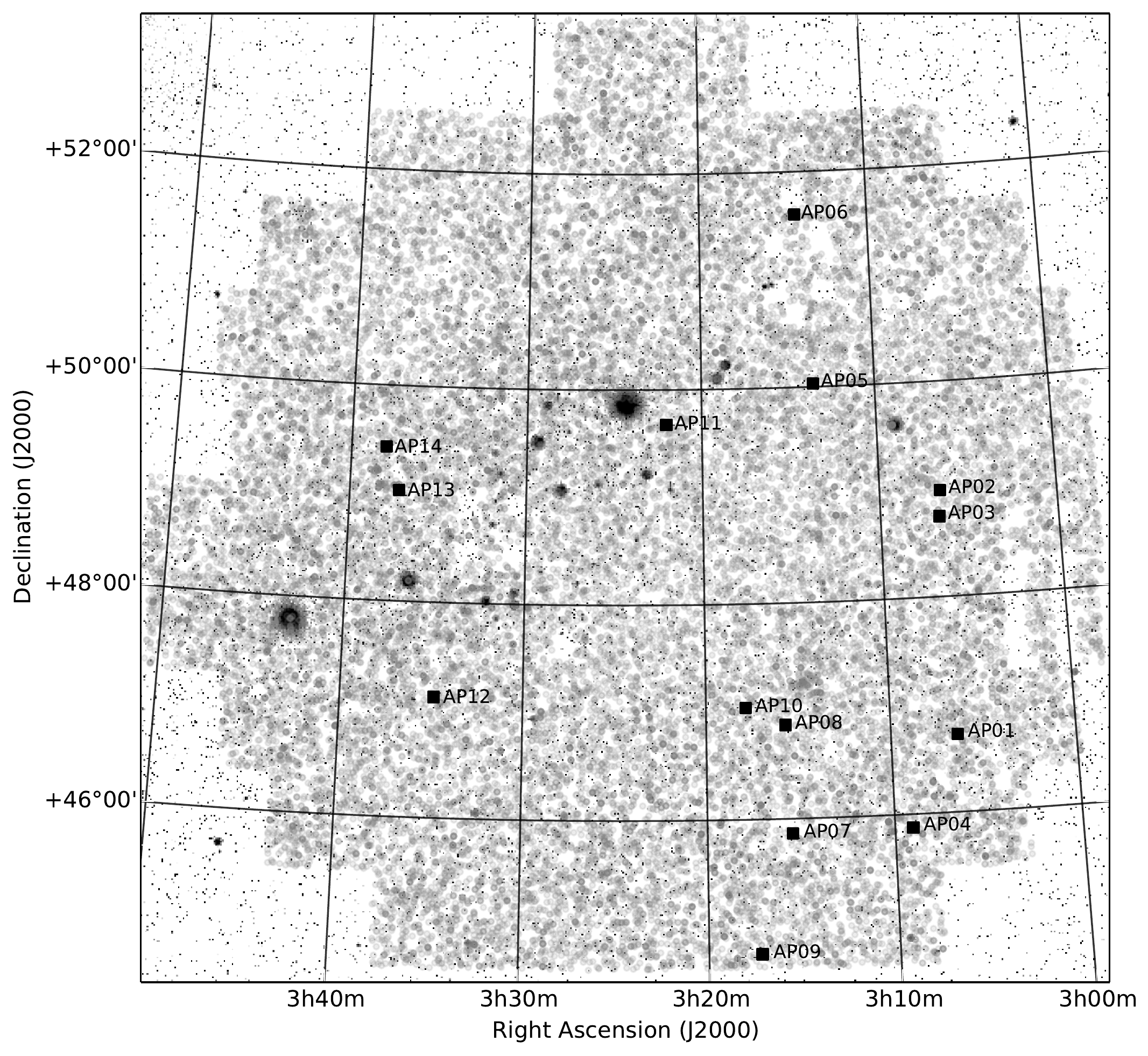}}
\caption{\label{map} The surveyed region of $\alpha$ Per }
\end{center}
\end{figure}

To predict the UKIDSS colours of likely $\alpha$ Per white dwarfs we used grids of model H-rich white dwarf  photometry appropriate to DA white dwarfs.  These grids are based upon the work of 
\citet{bergeron95} but are revised to include updates from \citet{holberg06}, \citet{kowalski06} and \citet{tremblay11}.  These model-based predictions and the cluster parameters, ($\tau$=120 Myr, m-M=6.27 and E$_{B-V}$=0.1 \citealt{vanleuwen09}) informed
our selection criteria.  We selected all objects (1) between RA of 02:48:00 and 03:52:00, and declination of 40 and 55 degrees from the UKIDSS GCS
(Figure \ref{map}; \citealt{lawrence07}); (2) with $Z > 15.0$, $J \leq 19.25$ (a conservative limit based on our assumed cluster parameters),  $0.1\geq Z-J \geq-0.5$
and $0.1 \geq Y-J \geq -0.5$; and (3) with classifiers in $Z$ and $Y$ = -1 (stellar) and the  post processing bit code to be less than 16, selecting the cleanest images \citep{hambly08}.  These criteria resulted in the selection of 2060 objects which were
then cross-matched with the SuperCOSMOS Sky Survey \citep{hambly01} within
2''. 

\begin{figure}
\begin{center}
\scalebox{0.25}{\includegraphics {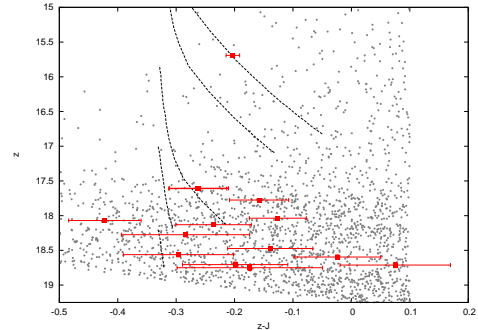}}
\caption{\label{cmd} $Z$, $Z-J$ colour magnitude diagram of the selected
  region. The UKIDSS selected objects are shown in grey, with the white dwarf
  candidates in red. The synthetic photometry from \citet{holberg06} is
  shown as dashed lines, each representing log g from 7.0 in steps of 0.5 to
  log g of 9.0, from brightest to faintest.}
\end{center}
\end{figure}

\begin{figure}
\begin{center}
\scalebox{0.3}{\includegraphics {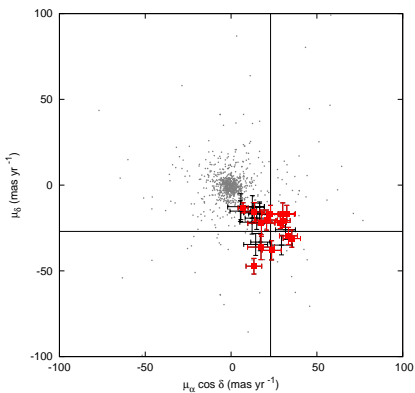}}
\caption{\label{pm} Proper motion diagram. The 2060 objects selected from the
  photometric selection using the colour-magnitude diagrams are shown in grey. The proper motion selected
  objects are shown in black with error bars, and the 14 white dwarf candidates
  are plotted as red boxes. with error bars. The cluster motion is marked at 23, -27 mas yr$^{-1}$.}
\end{center}
\end{figure}

The cross-matched sample was then further selected by proper
motion, $40.0>\mu_{\alpha}cos\delta >5.0$ mas yr$^{-1}$, $-10.0>\mu_{\delta}>-50.0$ mas yr$^{-1}$. This cut
removed objects that appeared to fall very close to the background object
centre of motion (0,0),  reducing the chances of contaminating objects being
selected. We then  requested that the errors on the proper motion be less than
8 mas yr$^{-1}$ and that the total proper motion be within 24 mas yr$^{-1}$ of
the cluster motion (23, -27 mas yr$^{-1}$; \citealt{vanleuwen09}). 
This cut resulted in 26 objects, which was thinned down to 14 after rejecting objects that were flagged as being blended in the SuperCOSMOS data. These objects can
be seen in Figures \ref{cmd} and \ref{pm} and in Table \ref{mags}.

\begin{table*}
\caption{\label{mags}ID, RA, dec, proper motion and UKIDSS magnitudes for the 14 $\alpha$ Per candidate white dwarfs.}
\begin{center}
\begin{tabular}{lccccccccc}
\hline
ID&RA&dec&$\mu_{\alpha}$Cos$\delta$&$\mu_{\delta}$&$Z$&$Y$&$J$&$H$&$K$\\

\hline
APWD01 & 03 06 24.00  & +46 43 11.3  &+28.72$\pm$3.49&$-$21.84$\pm$3.48&17.611$\pm$0.016& 17.849$\pm$0.026&17.874$\pm$0.047&18.013$\pm$0.077&18.085$\pm$0.127\\
APWD02 & 03 06 34.72  & +48 59 13.7  &+33.44$\pm$5.26&$-$29.61$\pm$4.99&18.130$\pm$0.024&18.267$\pm$0.028&18.366$\pm$0.060&18.197$\pm$0.087&18.801$\pm$0.228\\
APWD03 & 03 06 41.73  & +48 44 43.5  &+17.44$\pm$7.84&$-$36.12$\pm$7.33&18.712$\pm$0.037&18.682$\pm$0.0411&18.637$\pm$0.087&18.816$\pm$0.156&18.581$\pm$0.181\\
APWD04 & 03 09 02.57  & +45 52 34.2  &+22.98$\pm$4.98&$-$16.64$\pm$4.86&18.705$\pm$0.035&18.889$\pm$0.050&18.904$\pm$0.083&-&+19.513$\pm$0.389\\
APWD05 & 03 13 32.30  & +50 01 54.2  &+30.26$\pm$4.17&$-$20.65$\pm$3.96&17.605$\pm$0.021&17.828$\pm$0.029&17.866$\pm$0.046&17.731$\pm$0.080&17.962$\pm$0.107\\
APWD06 & 03 14 18.43  & +51 36 08.7  &+20.67$\pm$6.33&$-$20.38$\pm$5.89&18.750$\pm$0.040&18.801$\pm$0.060&18.924$\pm$0.118&-&-\\
APWD07 & 03 15 27.46  & +45 51 48.0  &+13.57$\pm$5.76&$-$15.68$\pm$5.51&18.076$\pm$0.022&18.333$\pm$0.032&18.492$\pm$0.058&-&-\\
APWD08 & 03 15 41.48  & +46 52 11.0  & +7.00$\pm$2.93&$-$13.24$\pm$2.96&15.692$\pm$0.005&15.883$\pm$0.007&15.895$\pm$0.0105&15.958$\pm$0.0173&16.025$\pm$0.026\\
APWD09 & 03 17 13.62  & +44 45 17.0  &+17.66$\pm$7.92&$-$22.18$\pm$7.66&18.037$\pm$0.021&18.177$\pm$0.029&18.163$\pm$0.044&18.124$\pm$0.089&18.636$\pm$0.263\\
APWD10 & 03 17 50.09  & +47 02 07.7  &+32.45$\pm$5.09&$-$16.76$\pm$4.97&18.274$\pm$0.032&18.564$\pm$0.052&18.558$\pm$0.105&18.426$\pm$0.150&-\\
APWD11 & 03 22 02.12  & +49 40 34.8  &+23.62$\pm$5.63&$-$38.04$\pm$5.62&17.775$\pm$0.019&17.961$\pm$0.030&17.933$\pm$0.047&17.950$\pm$0.086&17.972$\pm$0.112\\
APWD12 & 03 34 51.78  & +47 07 17.1  &+30.12$\pm$6.78&$-$17.03$\pm$6.70&18.473$\pm$0.029&18.528$\pm$0.038&18.612$\pm$0.067&18.445$\pm$0.138&18.688$\pm$0.213\\
APWD13 & 03 37 12.37  & +49 01 42.9  &+13.30$\pm$4.57&$-$47.27$\pm$4.57&18.562$\pm$0.031&18.746$\pm$0.047&18.858$\pm$0.089&-&-\\
APWD14 & 03 38 01.60  & +49 25 35.6  &+35.39$\pm$5.06&$-$31.18$\pm$5.06&18.596$\pm$0.0301&18.668$\pm$0.042&18.620$\pm$0.068&18.565$\pm$0.166&-\\
\hline
\end{tabular}
\end{center}
\end{table*}

Some estimates of the mass function suggest that $\alpha$ Per could be $\sim$2-3$\times$ as rich as the marginally older Pleiades which harbours at 
least one WD, LB1497 \citep{sanner01}. However, other mass functions suggest the two clusters are in fact very similar \citep{lodieu12}.
Thus assuming that the initial mass functions of these populations are at least comparable in form, it is possible that some of the white dwarf candidates listed in table \ref{mags} are 
 $\alpha$ Per members.

\section{Optical Spectroscopy}
To accurately measure the effective temperatures and surface gravities of the candidate white dwarf stars, as well as reliably assess cluster 
membership and estimate their initial and final masses, we obtained
medium resolution optical spectroscopy  of the 11 brightest candidates in Table \ref{mags}  using the Intermediate dispersion Spectrograph
and Imaging System (ISIS) on the William
Herschel Telescope on La Palma. We observed on the nights of 2011 September 03, 2011 September 04 and 2011 September 05 using the 300B  and 1200R gratings
simultaneously, with exposure times of between 2000 and 7200 s, split into at
least 2 separate exposures to aid with cosmic ray rejection. A 1" slit was used to provide spectral resolution of $\approx$3.5 \AA.

The spectra were reduced using \textsc{IRAF} \citep{tody1986, tody1993}. The CCD frames were debiased and flat fielded using \textsc{ccdproc} and cosmic ray hits were removed using \textsc{lacos spec} \citep{vandokkum01}. 
The spectra were then extracted using routines within the \textsc{apextract} package, and wavelength calibration was done using the CuAr+CuNe  arc spectra. We used the  white dwarf spectral standard stars GD71 \citep{greenstein69} and EG131 \citep{luyten49} to remove the instrument response and to provide
flux calibration.

On examining the spectra it became clear that only 7 of the white dwarfs, APWD01, APWD02, APWD04, APWD05, APWD07, APWD09 and APWD12 were DA white dwarfs, with APWD10, APWD11 and APWD13 appearing to be DBs. The remaining object APWD08, is a hot  subdwarf. 
%
\section{Modelling white dwarf spectra}
\label{dataanal}
The data were compared to the predictions of white dwarf model atmospheres using the spectral fitting programme FITSB2 (v2.04; \citealt{napiwotzki04}).  The same grid of pure-H model spectra was used as in \citet{casewell09}. It was calculated using the plane-parallel, hydrostatic, non-local thermodynamic equilibrium (non-LTE) atmosphere code 
TLUSTY, v200 \citep{hubeny88,hubeny95} and the spectral synthesis code SYNSPEC v48 \citep{hubeny01}. The models include a treatment for convective energy transport according to the ML2 prescription of \citet{bergeron92},
adopting a mixing length parameter, $\alpha$=0.6. These calculations utilised a model H-atom which incorporates explicitly the eight lowest 
energy levels and represents levels n=9 to 80 by a single superlevel. The dissolution of the high lying levels was treated by means of the occupation 
probability formalism of \citet{hummer88} generalised to the non-LTE atmosphere situation by \citet{hubeny94}. All calculations include the 
bound-free and free-free opacities of the H$^{-}$ ion and incorporate a full treatment for the blanketing effects of HI lines and the 
Lyman $-\alpha$, $-\beta$ and $-\gamma$ satellite opacities as computed by N. Allard \citep{allard04}. During the calculation of the model structure
the lines of the Lyman and Balmer series were treated by means of an Approximate Stark profile but in the spectral synthesis step detailed profiles 
for the Balmer lines were calculated from the Stark broadening tables of \citet{lemke97}. The grid of model spectra covered the T$_{\rm eff}$ range of 
13000-20000 K in steps of 1000 K and log g between 7.5 and 8.5 in steps of 0.1 dex. These models have been used throughout our work on the initial mass-final mass relation and 
we continue to use them in this work for consistency.

\begin{figure}
\begin{center}
\scalebox{0.3}{\includegraphics{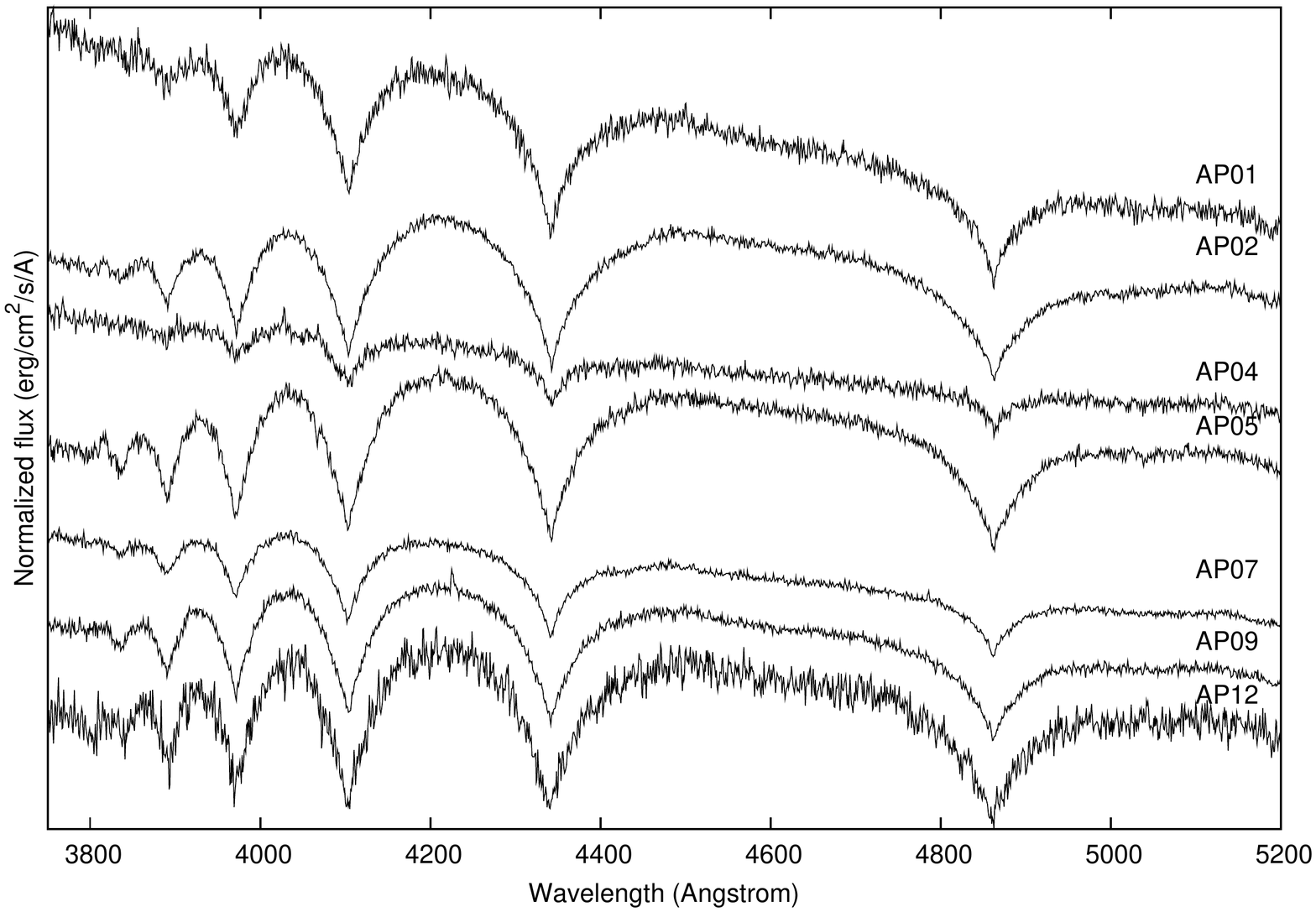}}
\caption{\label{da} ISIS spectra of the 7 DA white dwarfs. }
\end{center}
\end{figure}

 We used FITSB2 to fit our grid of model spectra to the DA white dwarf spectra using the seven Balmer 
absorption lines (H$\alpha$ is not in the observed wavelength range) ranging from H$\beta$ to H10. Points in the observed data lying more than 3$\sigma$ from the model were clipped from subsequent iterations of the fitting process (Figure \ref{da}). We list the results of model fitting for the DA white dwarfs in Table \ref{teff}.

\begin{figure}
\begin{center}
\scalebox{0.3}{\includegraphics{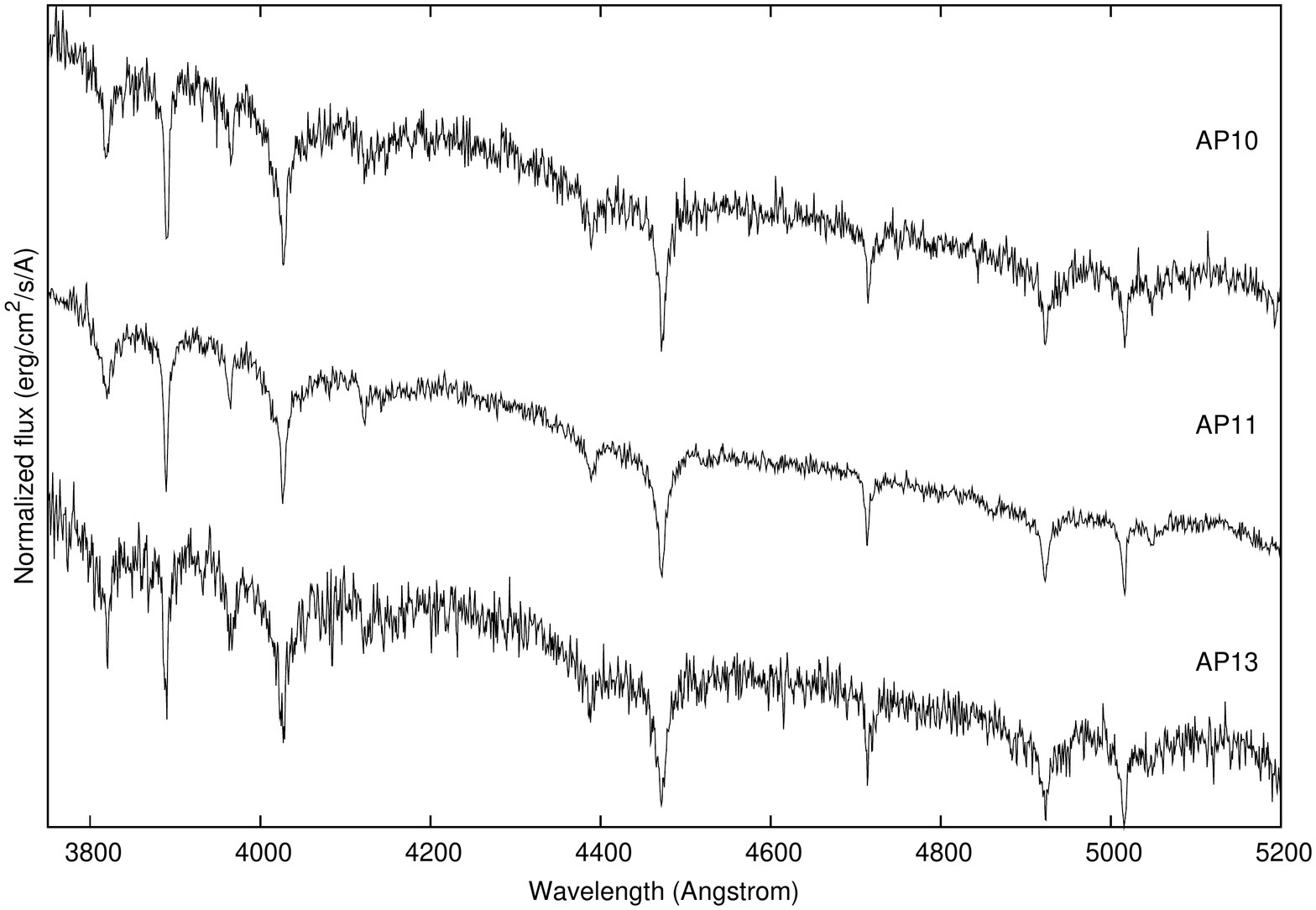}}
\caption{\label{db} ISIS spectra of the 3 DB white dwarfs}
\end{center}
\end{figure}

The effective temperature and surface gravity of the DBs was measured by comparing the normalised observed energy distribution with the wavelength range 3750-5150 \AA~to a grid of similarly normalised synthetic spectra. These were generated using \textsc{atm} and \textsc{syn}, tuned for the  treatment of helium rich atmospheres. The model fitting was undertaken using \textsc{xspec} as in \citet{baxter14}  and the results can be seen in Figure \ref{db}. 

\begin{figure}
\begin{center}
\scalebox{0.3}{\includegraphics{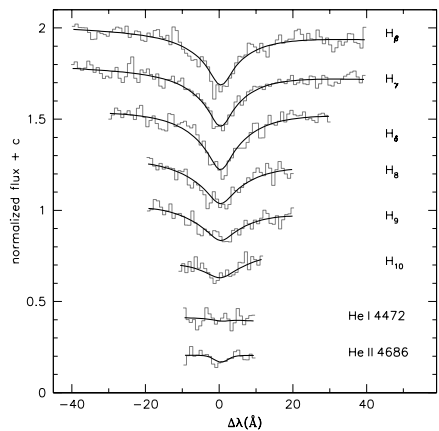}}
\caption{\label{sdb} T$_{\rm eff}$ and log g fit of APWD08, the sdOB star. The fit is the thick black line, and the data are shown as the lower resolution histogram.}
\end{center}
\end{figure}

The remaining object APWD08 was determined to be a hydrogen-rich sdOB star (Figure \ref{sdb}). The atmospheric parameters were derived as described in \citet{geier11} by fitting model spectra calculated in LTE and adopting a metal content of ten times the solar value to account for known peculiarities in hot subdwarf atmospheres caused by diffusion \citep{otoole06}.  Assuming the canonical sdB mass of 0.47 M$_{\odot}$, the distance to APWD08 is $\sim$ 2 kpc, and therefore it is definitely not a $\alpha$ Per  member.

\begin{table*}
\caption{\label{teff}ID,T$_{\rm eff}$, log g, and calculated mass and radii for the  11 observed $\alpha$ Per candidate white dwarfs.}
\begin{center}
\begin{tabular}{lccccccc}
\hline
ID&T$_{\rm eff}$ (K) & log g& Mass (M$_{\odot}$)&Radius (R$_{\odot}$)&Cooling age (Myr)\\
\hline
APWD01&26804$\pm$190&8.48$\pm$0.03&0.93$\pm$0.04& 0.938$\pm$0.084& 92.3$\pm$16.8\\
APWD02&14138$\pm$116&8.35$\pm$0.02&0.83$\pm$0.05& 1.034$\pm$0.093& 406.3$\pm$52.9\\ 

APWD04&35138$\pm$295&7.89$\pm$0.05&0.60$\pm$0.03& 1.498$\pm$0.133& 5.2$\pm$0.4\\
APWD05&15182$\pm$109&7.84$\pm$0.02&0.53$\pm$0.04& 1.493$\pm$0.133& 145.8$\pm$20.0\\ 
APWD07&24409$\pm$145&7.99$\pm$0.02&0.63$\pm$0.03& 1.356$\pm$0.085& 24.5$\pm$4.1\\
APWD08&38600$\pm$700&5.64$\pm$0.08&-&-&-\\
APWD09&12582$\pm$140&8.40$\pm$0.02&0.86$\pm$0.05& 0.993$\pm$0.090& 607.4$\pm$79.9\\
APWD10&16180$\pm$190&8.03$\pm$0.09&0.61$\pm$0.03& 1.280$\pm$0.080& 174.9$\pm$17.6\\
APWD11&15850$\pm$80&8.09$\pm$0.04&0.64$\pm$0.03& 1.227$\pm$0.077& 203.6$\pm$19.2\\
APWD12&13858$\pm$315&7.93$\pm$0.05&0.57$\pm$0.03& 1.388$\pm$0.087& 225.6$\pm$23.1\\

APWD13&16760$\pm$205&8.12$\pm$0.09& 0.66$\pm$0.03& 1.202$\pm$0.076& 177.7$\pm$17.5\\ 
\hline
\end{tabular}
\end{center}
\end{table*}

The errors given in Table \ref{teff} for T$_{\rm eff}$ and log g are formal fitting errors and are unrealistically small as they neglect systematic
uncertainties, e.g., flat fielding errors and model shortcomings. In subsequent discussion 
here we follow \citet{napiwotzki99} and assume an uncertainty of 2.3 per cent in  T$_{\rm eff}$ and 0.07 dex in log g for the DA white dwarfs, and an error of 2.0 per cent in T$_{\rm eff}$ and 0.05 dex in log g as in \citet{bergeron11} for the DB white dwarfs.  Both fitting programs also fit radial velocity. These were measured for the DA white dwarfs using H$\beta$ and H$\gamma$, and the He lines for the DB white dwarfs (Table \ref{teff}).

\section{Cluster membership of white dwarfs}

\begin{figure*}
\begin{center}
\scalebox{0.5}{\includegraphics[trim=1.8cm 0cm 0cm 0cm, clip=true, angle=0]{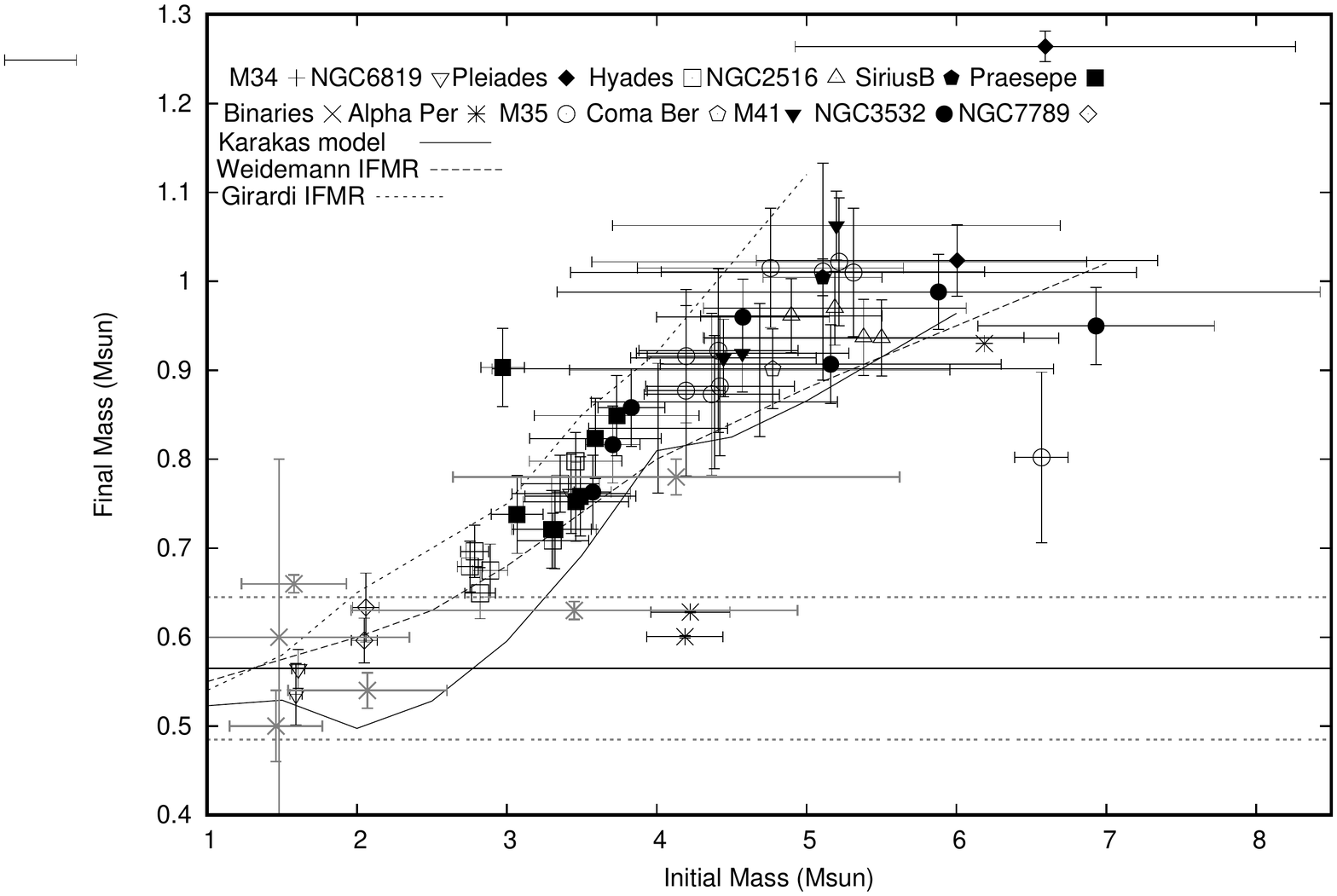}}
\caption{\label{ifmr} The IFMR of the available cluster and wide binaries data showing the position of the three $\alpha$ Per candidate members. The dashed black line is the semi-empirical \citet{weidemann00} IFMR, the thick solid line is the IFMR as given by the \citet{girardi00} models and the grey dot-dashed line is the initial mass-core mass at the first thermal pulse relation from \citet*{karakas02}. The peak in the field white dwarf mass distribution (thin solid line) and $\pm$1$\sigma$ is represented by the thin dotted lines. The plotted white dwarfs are from \citet{weidemann87,weidemann00, ferrario05, dobbie06, williams07, catalan08,kalirai08,rubin08, casewell09,dobbie09, williams09}.}
\end{center}
\end{figure*}

As in our earlier work (e.g. \citealt{casewell09}), we have utilised a grid of evolutionary models based on a mixed CO core 
composition and a thick H surface layer (e.g. \citealt{fontaine01}) to estimate the mass and cooling time 
of each DA white dwarf from our measurements of effective temperature and
surface gravity (see Table \ref{teff}). The same models, but with a thin H
surface layer were used for the DB white dwarfs. Cubic splines have 
been used to interpolate between the points within these grids. The lifetime of the progenitor star of each white dwarf has
then been calculated by subtracting the cooling time from the age of the cluster (90$\pm$10 Myr: \citealt{stauffer99}). 
Examining these cooling times, it is clear that only three objects have cooling times that are less than the cluster age, APWD01, APWD04 and APWD07.
The rest are much older, and are likely to be field objects. To constrain
the mass of the progenitor star for these three objects, we have used the stellar evolution models of \citet{girardi00} for solar metallicity, 
again using cubic splines to interpolate between the points in the grid. Assuming they are members of $\alpha$ Per, the progenitor mass for APWD01 is 6.185$\pm$0.402 M$_{\odot}$, APWD04 is 5.629$\pm$0.279M$_{\odot}$, and APWD07 is 5.734$\pm$0.305 M$_{\odot}$.  Both of the progenitor masses for APWD04 and APWD07 put these objects well below any semi-empirical and theoretical initial mass-final mass relation derived from cluster objects. (Figure \ref{ifmr}). APWD01 does however, look as though it may be a cluster member from its position on the diagram.

 The cluster radial velocity of $\alpha$ Per is -1.6 kms$^{-1}$ \citep{vanleuwen09} and using the mass and radius in Table \ref{teff}, the gravitational redshift and hence the absolute radial velocity of the candidate cluster members was determined (Figure \ref{rv}). 
It can be seen that none of the candidates have a radial velocity that is consistent with membership once the errors are taken into account, although APWD02 and APWD11 are closest to the cluster value. However, from Table \ref{teff}, the cooling age of these white dwarfs is much too long for them to be cluster members, and so we conclude that none of the white dwarfs studied here are members of the $\alpha$ Per open star cluster.
\begin{table}
\caption{\label{rv}Gravitational redshift and radial velocity for the 11 observed $\alpha$ Per candidate white dwarfs.}
\begin{center}
\begin{tabular}{lcccc}
\hline
ID&  V$_{\rm gr}$ (kms$^{-1}$) & RV (kms$^{-1}$)\\
\hline
 APWD01&   62.95$\pm$3.00 & $-$42.95$\pm$17.06\\  
 APWD02&   50.96$\pm$7.42 & 15.23 $\pm$10.91\\
 APWD04&   25.56$\pm$3.74 & 93.00$\pm$8.83\\
 APWD05&   22.37$\pm$3.55 & 15.09$\pm$8.75\\
 APWD07&   29.46$\pm$3.15 & $-$38.82$\pm$8.60\\
 APWD09&   54.98$\pm$7.96 & $-$57.92$\pm$11.28\\
APWD10&   30.36$\pm$3.34 & 49.94$\pm$8.67 \\
 APWD11&   33.46$\pm$3.65 & $-$12.16$\pm$8.79\\ 
 APWD12&   35.31$\pm$5.46 & $-$102.59$\pm$9.69\\
 APWD13&   35.18$\pm$3.83 & $-$20.18$\pm$8.87\\
\hline
\end{tabular}
\end{center}
\end{table}

%
%
\section{Discussion on the number of white dwarf members of $\alpha$ Per}

Our results were disappointing, as although we had a high level of success in identifying new white dwarfs, none are cluster members, despite the mass functions suggesting at least one white dwarf member is possible. As $\alpha$ Per is young, we do not expect it to contain many white dwarfs - the Pleiades only contains one, LB1497 \citep{eggen65}, and an additional object GD50, that may belong to the Pleiades moving group \citep{dobbie06}.  The fact that we have identified 11 white dwarf candidates, 10 of which are bona-fidae degenerates in the vicinity of $\alpha$ Per, and yet none are cluster members is perhaps highlighting the scarcity of white dwarfs in 100 Myr old clusters. 

There is also a known deficit of white dwarfs in open clusters \citep{weidemann92, vonhippel98}. This has been attributed in part to dynamical evolution causing white dwarfs to evaporate from clusters. Mass segregation alone is not sufficient to remove white dwarfs, as the average mass of a white dwarf ($\sim0.6 M_{\odot}$) is still more massive than most open cluster members and so is unlikely to suffer any effects to preferentially remove them from the cluster (e.g. \citealt{hurley03, baumgardt03}. It has however, been suggested that a velocity kick resulting from asymmetrical mass loss during post-main sequence evolution may preferentially remove white dwarfs from a cluster \citep{weidemann92, fellhauer03}. As $\alpha$ Per is younger than the Pleiades and so it is unlikely significant mass segregation has occurred within the cluster \citep{moraux03}. The answer to our question may simply be that $\alpha$ Per did not form many intermediate mass star white dwarf progenitors.

The age of $\alpha$ Per is 90 $\pm$10 Myr from measurements of the lithium depletion boundary \citep{stauffer99}. By using the models of \citet{girardi00} we determined the mass of a star that has a main sequence lifetime of 100 Myr to be 5.36 M$_{\odot}$. This is the minimum stellar mass required to form a white dwarf in $\alpha$ Per. We then combined this stellar mass with the IFMR presented in \citet{casewell09},  to give a white dwarf final mass of 1.0 M$_{\odot}$. This is the least massive white dwarf predicted to exist in the cluster. Such a white dwarf would have practically no cooling time.  

We used the same method to determine the maximum mass of a star that could form a white dwarf (assuming the maximum mass of a white dwarf is 1.4 M$_{\odot}$, which gives an initial mass of 8.71 M$_{\odot}$. Such a star has a main sequence lifetime of 4 Myr, and hence a maximum cooling age of 96 Myr. The models of \citet{fontaine01} give such a white dwarf a T$_{\rm eff}$ of 40000 K and a log g of 9.27, much higher than any of the white dwarfs detected in our survey. Indeed, even though the highest mass the \citet{holberg06} synthetic colours cover is 1.2 M$_{\odot}$, such a white dwarf should have a $J$ magnitude of 18.188 and $J-H$ =0.114 and $J-K$=-0.218.
Seven of the selected white dwarf candidates are fainter than this, however the majority have the correct colours suggesting that the colour cuts made were reasonable.  It may be that simply no white dwarfs have formed within the cluster age. This would appear to be borne out by the survey of \citet{heckmann58} who discovered 3 B3 stars, one of which is a subgiant within the cluster, and nothing of earlier spectral types.
These data were used by \citet{lodieu12} who created a detailed mass function of the cluster, but again, the highest mass stars discovered in the cluster to date are 5 M$_{\odot}$, and there are 8$\pm$3 stars in this mass bin which ranges from 4.13 to 5 M$_{\odot}$. Extrapolating this relationship into the region we are interested in gives $\sim$1.4 high mass stars at 5.36 M$_{\odot}$ and only $\sim$0.4 at 8.71 M$_{\odot}$ assuming a bin size of 1 M$_{\odot}$, again confirming the hypothesis that there are very few, if any, high enough mass stars in $\alpha$ Per to form high mass white dwarfs at this young age.

Another possible scenario is that the white dwarfs remain bound to the cluster, but are hidden in binary systems, as for the Hyades \citep{bohm93, franz98, debernardi00}.  \citet{williams04b} simulated three clusters to investigate the initial mass functions and the probability of hidden high mass white dwarfs in binaries. They found that for the Pleiades, their simulations agreed with the observations and that LB1497 is likely to be the only white dwarf in the cluster. Their work on Praesepe and the Hyades however, highlighted the lack of high mass white dwarfs, and they suggest that it is more likely for high mass progenitor stars to be located in binaries with massive evolved stars, than with lower mass companions, thus making them less likely to be detected. It may be that any high mass white dwarf members of the cluster are in such binaries, but again, the lack of high mass stars in general in $\alpha$ Per makes it seem unlikely.

The final scenario to consider is that the white dwarf candidates in this work were selected for their high reliability of being white dwarfs, which was prioritised over the completeness of the survey. The UKIDSS GCS is estimated as being 100 per cent complete over the magnitude range we studied \citep{lodieu12}, and the image morphology selection of class =-1, ensures only stellar-like sources are selected, although this is known to be conservative. The SuperCOSMOS data however, was cross-matched to this clean GCS sample, and even using the COSMOS crowded field algorithm, it is estimated that the completeness is only 60 per cent at this low Galactic latitude \citep{beard90}.  This low completeness will dominate our survey and may mean we have potentially missed the one or two expected white dwarf cluster members. The data from Gaia should locate any missing white dwarfs in this cluster, should they be there.

\section{Summary}
We have obtained spectra of 11 white dwarf candidate members of the $\alpha$ Per open star cluster, and confirmed that while 7 are DA white dwarfs, 3 are DB white dwarfs and one is an sdOB star, none are cluster members. This result is disappointing, as recent work on the cluster mass function suggests that there should be at least one white dwarf member, even at this young age. It may be that any white dwarf members of $\alpha$ Per are hidden within binary systems, as is the case in the Hyades cluster, however the lack of high mass stars within the cluster also makes this seem unlikely. We also conclude that the high  level of incompleteness in the SuperCOSMOS survey may mean that we have missed any white dwarf members, although Gaia is likely to locate them, should they exist.

%
\section*{Acknowledgements}
SLC acknowledges the support of the College of Science and Engineering at the University of Leicester. NL was funded by the Ram\'on y Cajal fellowship number 08-303-01-02. 
Based on observations made with the William Herschel Telescope operated on the island of La Palma by the Isaac Newton Group in the Spanish Observatorio del Roque de los Muchachos of the Instituto de Astrof\'isica de Canarias. The UKIDSS project is defined in \citet{lawrence07}. UKIDSS uses the UKIRT Wide Field Camera (WFCAM; \citealt{casali07} and a photometric system described in \citet{hewett06}. The pipeline processing and science archive are described in Irwin et al (in prep) and \citet{hambly08}. This research has made use of data obtained from the SuperCOSMOS Science Archive, prepared and hosted by the Wide Field Astronomy Unit, Institute for Astronomy, University of Edinburgh, which is funded by the UK Science and Technology Facilities Council. This research has also made use of NASA's Astrophysics Data System. 
%
%

\bibliographystyle{mn2e}
%

\bibliography{mnemonic,refs}
\label{lastpage}

\end{document}